\documentclass[twocolumn,secnumarabic,amssymb,amsfonts,nobibnotes,aps,jcp,showpacs]
{revtex4-1}

\usepackage{amsmath,amssymb,mathtools,mathrsfs}
\usepackage{latexsym}
\usepackage{float}
\usepackage{amsfonts}
\usepackage{graphicx}
\usepackage{textcomp}
\usepackage{hyperref}
\usepackage {pifont}

 1

\newcommand{\dummy}

\usepackage[usenames,dvipsnames]{xcolor}

\begin{document}
\title{Study of Critical Dynamics in Fluids via Molecular Dynamics in Canonical Ensemble}
\author{Sutapa Roy and Subir K. Das$^*$}
\affiliation{Theoretical Sciences Unit, Jawaharlal Nehru 
Centre for Advanced Scientific Research, Jakkur P.O, 
Bangalore 560064, India\\}
\date{\today}

\begin{abstract}
With the objective of understanding the usefulness of thermostats in the 
study of dynamic critical phenomena in fluids, we present results for 
transport properties in a binary Lennard-Jones fluid that exhibits 
liquid-liquid phase transition. Various collective transport 
properties, calculated from the molecular dynamics (MD) simulations in canonical 
ensemble, with different thermostats, are compared with those obtained from 
MD simulations in microcanonical ensemble. It is observed that the Nos\'{e}-Hoover and 
dissipative particle dynamics thermostats are useful for the calculations 
of mutual diffusivity and shear viscosity. The Nos\'{e}-Hoover thermostat, 
however, as opposed to the latter, appears inadequate for the study of bulk viscosity.
\end{abstract}
\maketitle

\section{Introduction}\label{introduction}
\par
\hspace{0.2cm}In the vicinity of a critical point \cite{zinn}, various static 
\cite{zinn,stanley,onuki1} and dynamic \cite{onuki1,onuki2,hohenberg,ferrell1,
olchowy,luettmer,anisimov,burstyn1,burstyn2} quantities exhibit power-law 
singularities. Computer simulations played a crucial role in the understanding 
of static critical phenomena \cite{landau}. In dynamics, on the other hand, 
simulations are recent, particularly for fluid criticality. In this case, in addition 
to the finite-size effects, critical slowing down poses enormous difficulty. 
Note that the slowest relaxation time, $\tau_{_{max}}$, diverges at the criticality as 
\cite{onuki1,landau}
\begin{eqnarray}\label {tau}
\tau_{_{max}} \sim L^z,
\end{eqnarray}
where $L$ is the linear dimension of the system and $z$ is a dynamic critical 
exponent. While in the static 
critical phenomena, the problem of critical slowing down can be significantly 
reduced via a smart choice of ensemble (with smaller value of $z$ ), in dynamics 
this is not possible. The liberty in statics 
stems from the robust universality of static critical 
phenomena.

\par
\hspace{0.2cm}For the computational study of critical dynamics in fluids, using microscopic 
models, one typically carries out molecular dynamics (MD) \cite{frenkel,allen,rapaport} 
simulations. Usually one considers the microcanonical ensemble (constant N, V, E, 
which are respectively the total number of particles, confining volume and energy) 
where requirements of hydrodynamics are satisfied. However, as seen in 
Eq. (\ref{tau}), close to the critical point, overwhelmingly long simulation runs 
are required to avoid finite-size effects even at a moderate level. In such a 
situation, control of temperature ($T$) in the NVE ensemble becomes problematic. A 
representative case for temperature drift in microcanonical runs has been shown 
in Fig. 1. Drift of such magnitude is acceptable in the normal region of the 
parameter space, i.e., far away from any phase transition. However, close to the critical point, where one focuses on 
quantifying singular behavior, this becomes problematic. This calls for the 
study of fluid critical dynamics in canonical (NVT) 
ensemble where, instead of $E$, $T$ is kept constant. 

\par
\hspace{0.2cm}Various thermostats \cite{frenkel,allen} are used to maintain 
temperature in MD simulations in NVT ensemble, e.g. Andersen thermostat (AT), 
Langevin thermostat (LT), Nos\'{e}-Hoover thermostat (NHT), dissipative particle 
dynamics thermostat (DPDT), etc. Even though most of the thermostats are useful 
in controlling the temperature of the system, only a few are appropriate for the
calculation of transport properties in fluids. Crucial tests of a 
thermostat, in terms of providing the correct value of a transport quantity as 
well as in keeping the temperature constant, lie in nontrivial phenomena like 
phase transition dynamics. In a recent work 
\cite{roy2014}, we have demonstrated the usefulness of the NHT for the calculation of 
shear viscosity. In this paper we address this issue in a more general context.

\par
\hspace{0.2cm}In AT \cite{frenkel}, the temperature is controlled via the random 
assignments of velocities to a fraction of particles according to the Maxwell 
distribution, mimicking collisions of the particles with a heat bath. Due to this 
Monte Carlo-like stochastic nature, AT is not useful for the calculation of 
transport properties in fluids. With increasing collision frequency, the transport coefficients 
deviate further and further from the desired value. This stochastic character is 
also true for LT. 

\par
\hspace{0.2cm}
For MD in NVE ensemble, one solves the Newton's equations 
of motion involving the inter-particle force. Like AT, in the NVT ensemble, depending 
upon the thermostat, additional rules are imposed. In the case of LT, an additional drag 
force proportional to the velocity is introduced, in addition to a random force, both 
coming from the background solvent particles. There, for the $i$th particle, one solves the 
equation \cite{grest}
\begin{eqnarray}\label {LT1}
\frac{d^2 {\vec r}_i}{dt^2}=-\vec {\nabla} U_i-
\gamma \frac{d{\vec r}_i}{dt}+{\vec W}_i,
\end{eqnarray}
where ${\vec r}_i$ is the position of the particle, $U_i$ is the inter-particle potential, $t$ is the 
time, $\gamma$ the drag coefficient and ${\vec W}_i$ is a temperature-dependent 
Gaussian noise with mean zero. The noise correlation between two times $t$ and $t'$ follows the 
fluctuation-dissipation relation
\begin{eqnarray}\label {LT2}
\langle {W_{i\mu}W_{j\nu}} \rangle=2k_BT{\gamma}\delta_{ij}
\delta(t-t')\delta_{\mu\nu}.
\end{eqnarray}
In Eq. (\ref{LT2}), $\mu$ and $\nu$ correspond to the Cartesian axes of space coordinates 
and $k_B$ is the Boltzmann constant. In case of non-Gaussian noise, one needs to 
appropriately adjust the numerical factor in Eq. (\ref{LT2}). In this work we have used 
uniform random numbers between $-1$ and $1$, thus the prefactor $2$ is replaced by $6$. 

\par
\hspace{0.2cm}
Due to their inability to conserve the local momentum, AT and LT are used only
for the equilibration purpose. Nevertheless, for the sake of completeness, we will present 
some results using these thermostats as well. But, there exist a number of thermostats, 
e.g. NHT, DPDT, etc., that are believed to be good for the calculation of transport 
properties in fluids. The understanding of the usefulness of these thermostats, however, 
to the best of our knowledge, is essentially restricted to the single particle dynamics.

\par
\hspace{0.2cm}In DPDT \cite{stoyanov,nikunen,soddemann}, the dissipative force in Eq. 
(\ref{LT1}) is given by $\gamma \omega^D(r_{ij})({\vec v}_{ij} \cdot {\vec e}_{ij}){\vec e}_{ij}$ 
where ${\vec r}_{ij}$ and ${\vec v}_{ij}$ are respectively the relative position and 
velocity between $i$ and $j$ particles with ${\vec e}_{ij}={\vec r}_{ij}/{r_{ij}}$; 
$r_{ij}=|{\vec r}_{ij}|=r$. Here, $\omega^D$ is a weight function connected to the random force as 
$\sqrt{2\gamma k_B T \omega^D}\omega_{ij}{\vec e}_{ij}$, where $\omega_{ij}$ are random 
numbers with $\omega_{ij}=\omega_{ji}$. For the choice of $\omega^D$, there is no 
fixed prescribed rule. In this work we use \cite{pastorino} $\omega^D=(1-r)^2$ for 
$r\le 1$ and $0$ otherwise. From the property of the random force and the expression of the 
dissipative force, it is understandable that DPDT will preserve local momentum, thus 
hydrodynamics. However, this thermostat has issues related to keeping the temperature 
constant. For the choice of the weight function mentioned above and $\gamma=0.1$, we 
obtained reasonable temperature control in this work. Note that for LT we used $\gamma=1$.

\par
\hspace{0.2cm}
In NHT, an additional degree of freedom $\Xi$ is introduced and one solves the equations 
\cite{frenkel}
\begin{eqnarray}\label {NHT1}
m_i \dot {\vec {r}}_i=\vec {p}_i,
\end{eqnarray}
\begin{eqnarray}\label {NHT2}
\dot {\vec {p}}_i=-\frac{\delta U_i}{\delta \vec {r}_i}-\Xi \vec {p}_i, 
\end{eqnarray}
\begin{eqnarray}\label {NHT3}
\dot \Xi=\Big(\sum_{i=1}^N p_i^2/m_i-3N/\beta \Big)/Q,
\end{eqnarray}
where $\beta=1/{k_BT}$, $\Xi$ is a time dependent drag, $\vec {p}_i$ is the momentum and 
$Q$ is the coupling strength between the system and the thermostat. Essentially, in this case, 
the simulation is done in microcanonical ensemble \cite{frenkel,stoyanov} with a modified 
Hamiltonian that provides averages equivalent to those of a canonical ensemble with the original 
Hamiltonian. The original energy function, that is constant in microcanonical
ensemble, fluctuates in this method, as in the
canonical ensemble. The constant of motion here is related to the Helmholtz free energy.
Unless otherwise mentioned, for all our presented results the value of $Q$ was
set to unity.

\par
\hspace{0.2cm}
As is clear by now, in this paper we provide results for the utility of NHT and DPDT with respect to the 
study of dynamic critical phenomena. Despite its problems related to local 
momentum conservation, NHT still remained popular for the study of 
transport using NVT ensemble. Of course, every hydrodynamic preserving thermostat has some 
disadvantages, e.g., DPDT suffers from the temperature control problem.

\par
\hspace{0.2cm}The rest of the paper is organized as follows. In Section II, we introduce 
the model. The results are presented in Section III. Finally, the paper is concluded in 
Section IV with a summary and discussion.

\section{Model and Phase Behavior}\label{model}
\par
\hspace{0.2cm}In our binary ($A+B$) mixture model \cite{das3,das4,royepl}, particles 
interact via the Lennard-Jones (LJ) pair potential
\begin {eqnarray}\label{LJ1}
u(r)=4\varepsilon_{_{\alpha\beta}}\Big[\Big(\frac{\sigma}{r}\Big)^{12}
-\Big(\frac{\sigma}{r}\Big)^{6}\Big],
\end{eqnarray}
where $\sigma$ is the particle diameter and $\varepsilon_{\alpha\beta}~[\alpha,\beta=A,B]$ 
is the interaction strength. For the sake of computational convenience, we have 
introduced a cut-off and shifting of the potential to zero at $r_c=2.5\sigma$. This, 
however, introduces a discontinuity in the force at $r_c$, which was removed by adding a term \cite{allen} 
$(r-r_{_c}) ({du}/{dr})_{{r}=r_{_c}}$. We work with a symmetric model by setting 
$\varepsilon_{_{AA}}=\varepsilon_{_{BB}}=2\varepsilon_{_{AB}}=\varepsilon$ which produces 
liquid-liquid phase separation. The overall density of particles was set to unity. This 
avoids overlap between liquid-liquid and vapor-liquid phase separation.

\par
\hspace{0.2cm}The phase diagram for this model was studied \cite{das3,das4,royepl} via a semi 
grandcanonical Monte Carlo \cite{landau,frenkel} method. In this scheme, in addition to 
the standard particle displacement moves, one tries identity switches 
$A \rightarrow B \rightarrow A $ which are accepted or rejected according to the standard 
Metropolis criterion. For the identity moves it is necessary to include in the Boltzmann 
factor \cite{frenkel} the chemical potential difference between the two species. This 
difference, however, is zero along 
coexistence and for $50:50$ composition above the critical temperature $T_c$, due to the 
symmetry of the model. In this 
ensemble, from the fluctuation of concentration $x_\alpha(=N_\alpha/N,$ $N_\alpha$ being 
the number of particles of species $\alpha$), one obtains a probability distribution 
$P(x_\alpha)$. Below the critical temperature, $P(x_\alpha)$ should have a two-peak 
structure, the locations of the peaks providing the points along the coexistence. At the critical 
temperature, the form of the distribution crosses over from the double peak to a single peak one. 
But, this critical temperature is system-size dependent that we will denote as $T_c^L$, 
which, in the limit $L \rightarrow \infty$, will converge to the thermodynamic critical 
temperature, $T_c$. In Table I we list the values of $T_c^L$ for a few system sizes 
\cite{royepl,royjcp}. 

\par
\hspace{0.2cm}
As already mentioned, transport properties are studied via MD simulations in NVE as well as NVT ensembles, 
for the latter various temperature controlling methods, discussed in the previous section, 
were used. Details on the calculation of transport properties will be provided in the 
next section.

\par
\hspace{0.2cm}All our simulations were performed in three space dimensions with cubic boxes 
of linear size $L$ (in units of $\sigma$) and periodic boundary conditions in all directions. The equations of 
motion in MD were solved by applying Verlet velocity algorithm with integration time step 
$\Delta t=0.005$. Before starting the production runs, the configurations were equilibrated 
via MC simulations and, in the case of MD in NVE ensemble, further thermalization runs were 
performed via MD with AT. Except for self diffusivity, results are presented after averaging 
over a very large number of independent initial configurations, ranging between $80$ and $640$. 
In case of self-diffusivity, this number is $5$. For collective properties, as the 
terminology suggests, such high numbers become necessary due to lack of averaging 
involving the individual particles. 

\begin{table}
\caption {} \label{table1} 
\begin{center}
  \begin{tabular}{|c|c|c|c|c|c|c|}
    \hline
    $L$ & 8 & 10 & 12 & 14 & 16 & $\infty$ \\ \hline
    $T_c^L$ & 1.461 & 1.447 & 1.440 & 1.436 & 1.433 & $T_c \simeq 1.421$ \\
    \hline
  \end{tabular}
\end{center}
\end{table}

\section{Results}\label{result}

\par
\hspace{0.2cm}Using MD, at various temperatures (fixing the composition to the critical value) we 
present results for the self diffusivity ($D$), Onsager coefficient ($\mathscr L$), shear 
viscosity ($\eta$) and bulk viscosity ($\zeta$). These quantities were calculated (in dimensionless units) 
from the Green-Kubo (GK) relations \cite{hansen} as (note that, because of the symmetry of the 
model $D=D_A=D_B$, $D_\alpha$ being the self diffusivity of species $\alpha$)
\begin{eqnarray}\label{da}
D(t)=\Big({\frac{t_0}{3\sigma^2}}\Big)\int_0^t dt' 
\langle {{\vec v}_{i,\alpha}(t'){\vec v}_{i,\alpha}(0)}\rangle,
\end{eqnarray}

\begin{eqnarray}\label{ons1}
{\mathscr L}(t)=\left(\frac {t_{_0}\varepsilon}{3k_{_B}NT\sigma^2}\right) 
\int_0^t dt' \langle {{\vec J}_{_{AB}}}(t'){{\vec J}_{_{AB}}}(0) \rangle,
\end{eqnarray}
 
\begin{eqnarray}\label{shear1}
\eta(t)={\left(\frac {t_{_0}^3 \varepsilon}{\sigma VTm^2}\right)
\int_0^t dt' 
\langle {P_{\mu\nu}}(t'){P_{\mu\nu}}(0)\rangle},
\end{eqnarray}
and
\begin{eqnarray}\label{bulk1}
Y(t)={\left(\frac {t_{_0}^3 \varepsilon}{\sigma VTm^2}\right)
\int_0^t dt' 
\langle {P^{'}_{\mu\mu}}(t'){P^{'}_{\mu\mu}}(0)\rangle},
\end{eqnarray}
where $t_0$ is the LJ time unit $(=\sqrt{m\sigma^2/\varepsilon})$ and $m$ is the mass 
(same for all particles in our model). In Eq. (\ref{ons1}), ${\vec J}_{AB}$ is a 
concentration current defined as
\begin{eqnarray}\label{ons2}
{\vec J}_{_{AB}}(t)={{x_{_B}}{\sum}_{i=1}^{N_{_A}}{\vec v_{_{i,A}}}(t)}
-{{x_{_A}}{\sum}_{i=1}^{N_{_B}}{\vec v_{_{i,B}}}(t)},~~
\end{eqnarray}
${\vec v}_{_{i,\alpha}}$ being the velocity of $i$th particle of species $\alpha$. In 
Eq. (\ref{shear1}), $P_{\mu\nu}$ are the off-diagonal elements of the pressure tensor 
given as 
\cite{hansen}
\begin{eqnarray}\label{ons3}
P_{\mu\nu}(t)=\sum_{i=1}^N\Big[mv_{i\mu}v_{i\nu}+
\frac{1}{2}\sum_{j(\ne i)}(\mu_i-\mu_j)F_\nu(|{\vec r}_i-{\vec r}_j|)\Big],~~~~~
\end{eqnarray}
$\vec F$ being the force between particles $i$ and $j$; $\mu_i$ is a Cartesian coordinate 
for the position of particle $i$. In Eq. (\ref {bulk1}), $Y=\zeta+\frac{4}{3} \eta$ and 
$P^{'}_{\mu\mu}=P_{\mu\mu}-P$, $P$ being the pressure. 

\par
\hspace{0.2cm}
These quantities can also be 
calculated from the corresponding mean squared displacements (MSD) following the Einstein 
relations, e.g., the self diffusivity $D$, the Onsager coefficient ${\mathscr L}$ and the shear 
viscosity $\eta$ are calculated as \cite{hansen}

\begin{eqnarray}\label{selfd}
D(t)=\Big(\frac{t_0}{6t\sigma^2}\Big)
\langle |{\vec r}_{i,\alpha}(t)-{\vec r}_{i,\alpha}(0)|^2 \rangle,
\end{eqnarray}
and

\begin{eqnarray}\label{ons10}
{\mathscr L}(t)=\left(\frac{{t_{_0}}N_A^2\varepsilon}
{6k_{_B}tNT\sigma^2}\right)
\langle |{\vec R}_{_A}(t)-{\vec R}_{_A}(0)|^2\rangle,
\end{eqnarray}
and 
\begin{eqnarray}\label{shearmsd}
\eta(t)=\left(\frac {t_{_0}^3\varepsilon}{2k_{_B}tVd_0T m^2}\right)
\langle |Q_{_{xy}}(t)-Q_{_{xy}}(0)|^2\rangle.
\end{eqnarray}
In Eq. (\ref{ons10}), ${\vec R}_{\alpha}$ is the centre of mass (CM) coordinate of species $\alpha$ 
and in Eq. (\ref{shearmsd}), the generalized displacement $Q_{xy}$ has the expression
\begin{eqnarray}\label{shearmsd2}
Q_{_{xy}}(t)=\sum_{i=1}^N {x_{_i}(t)v_{iy}(t)}.
\end{eqnarray}

In the rest of the paper, we set $m$, $\varepsilon$, $\sigma$, $t_0$ and $k_B$ to unity. For 
self diffusivity, Onsager coefficient and shear viscosity, we present results from the MSD relations whereas the 
results for bulk viscosity were obtained using the GK relation.

\begin{figure}[htb]
\centering
\includegraphics*[width=0.4\textwidth]{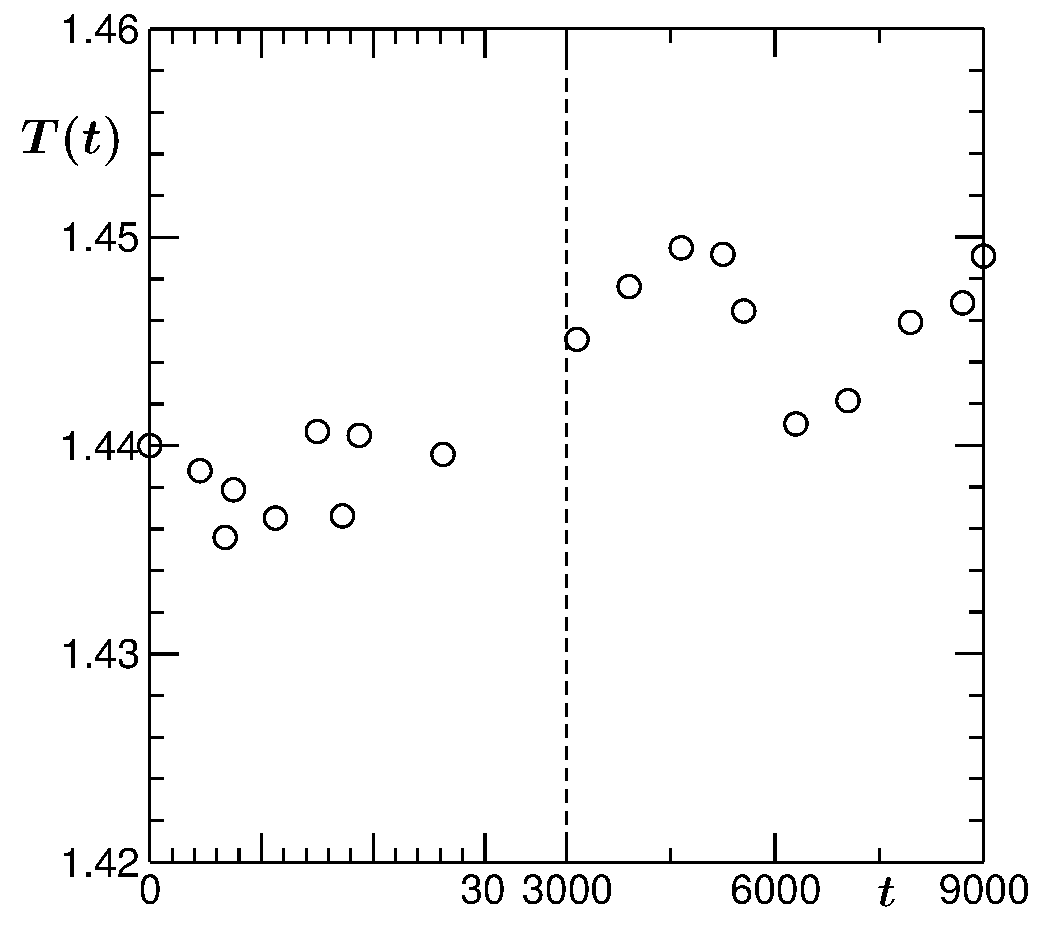}
\caption{\label{fig1}Drift of temperature is demonstrated for a typical molecular dynamics 
run in microcanonincal ensemble. The chosen temperature is very close to the critical value. 
The cubic simulation box has a linear dimension $L=14$ and number 
of particles $2744$. The initial temperature is set to a value $1.44$.}
\end{figure}

\begin{figure}[htb]
\centering
\includegraphics*[width=0.4\textwidth]{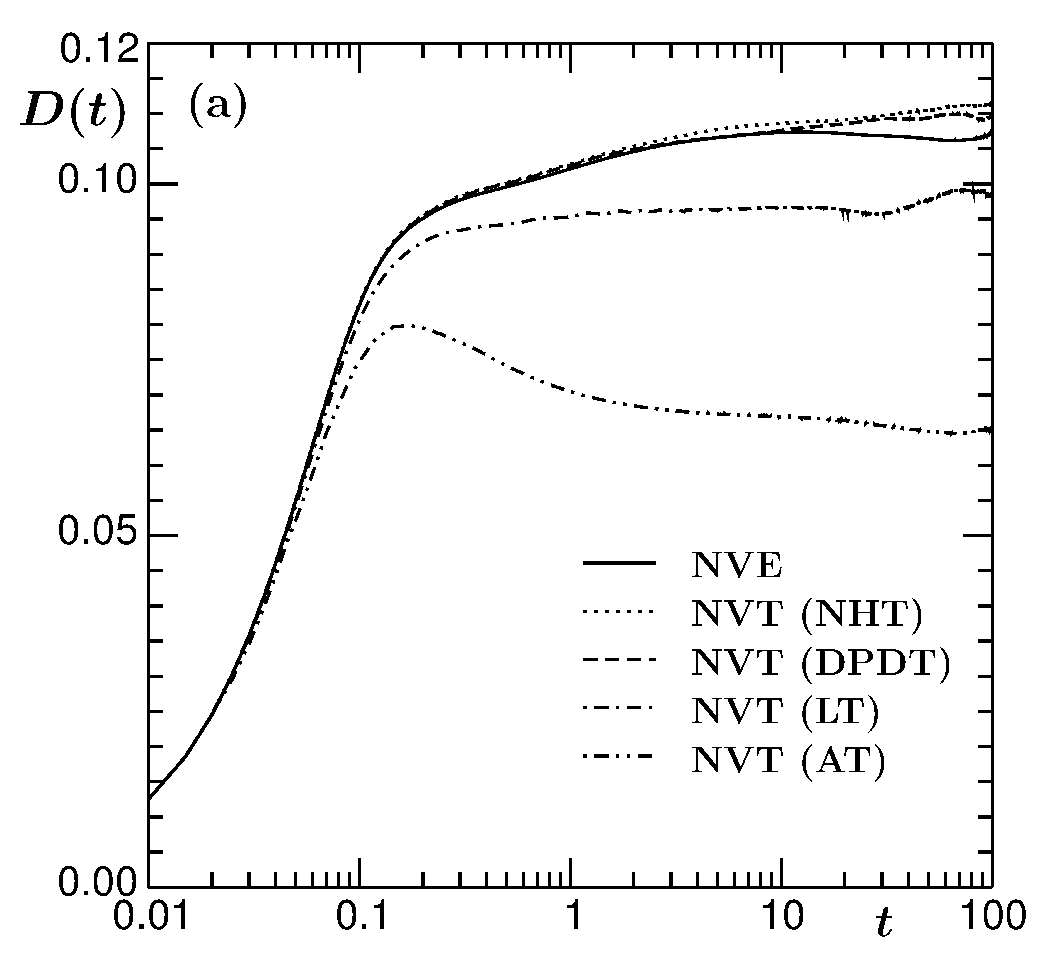}
\vskip 0.5cm
\includegraphics*[width=0.4\textwidth]{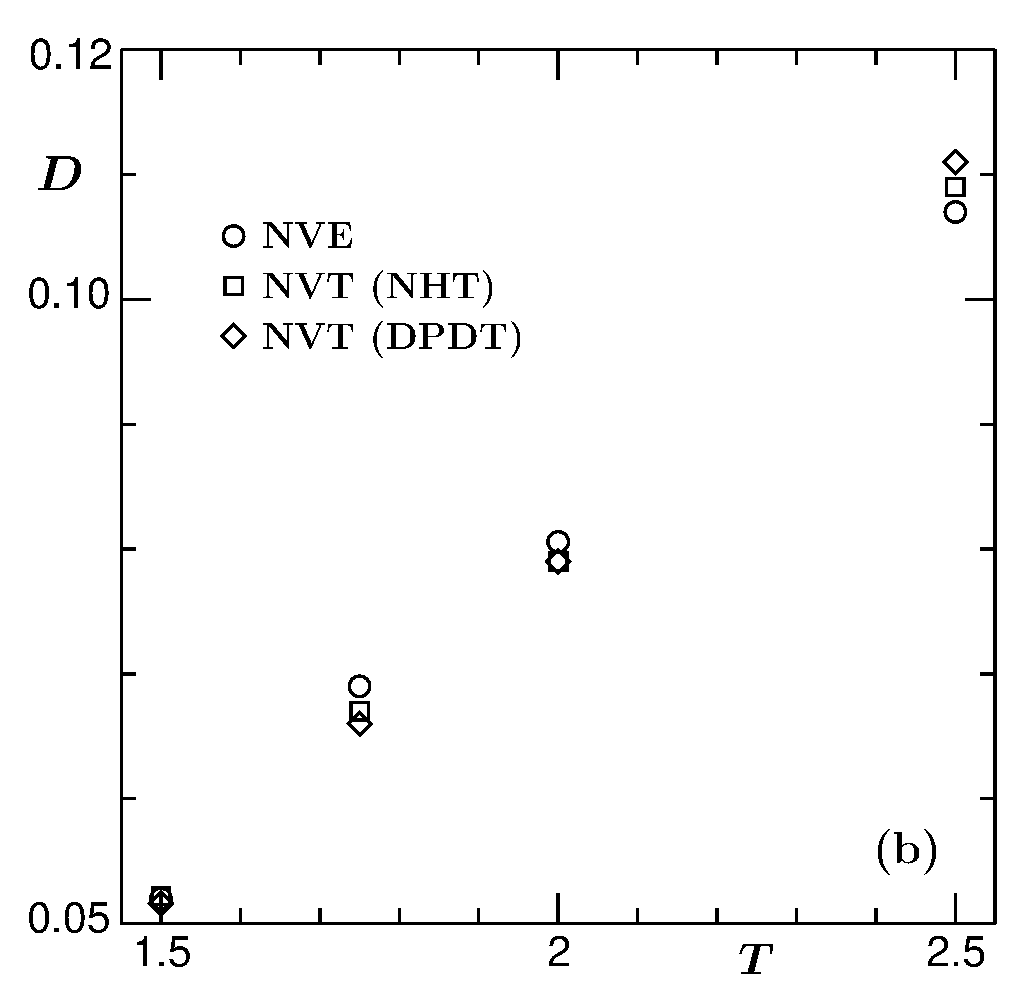}
\caption{\label{fig2}(a) Plot of time dependent self diffusivity obtained using various 
ensembles. (b) Plot of $D~[D(\infty)]$ vs $T$, for NVE, NHT and DPDT. All results are obtained 
after averaging over $5$ independent initial configurations.}
\end{figure}

\begin{figure}[htb]
\centering
\includegraphics*[width=0.4\textwidth]{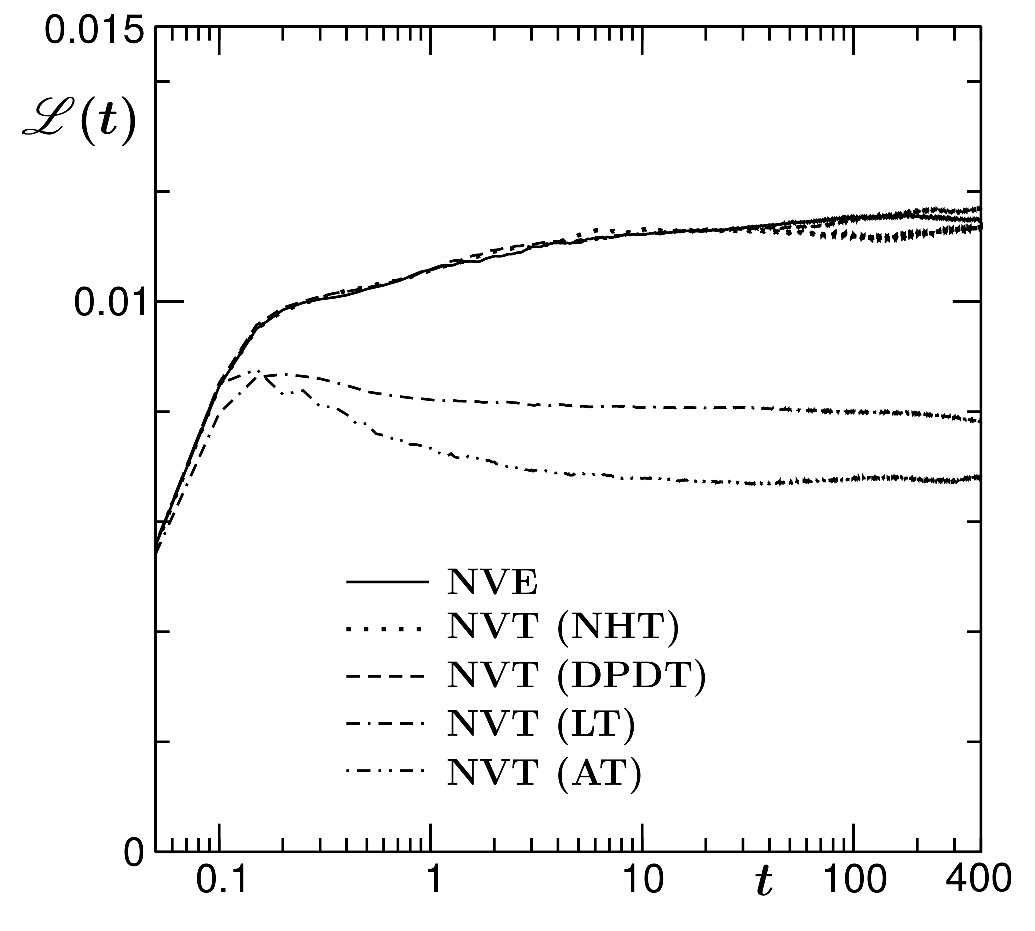}
\caption{\label{fig3}Plot of Onsager coefficient as a function of time, from MD calculations 
in NVE and NVT ensembles. For NVT ensemble, as indicated, four different thermostats were used. 
In all the cases, values of T and L were set to $2.5$ and $10$, respectively.}
\end{figure}

\begin{figure}[htb]
\centering
\includegraphics*[width=0.4\textwidth]{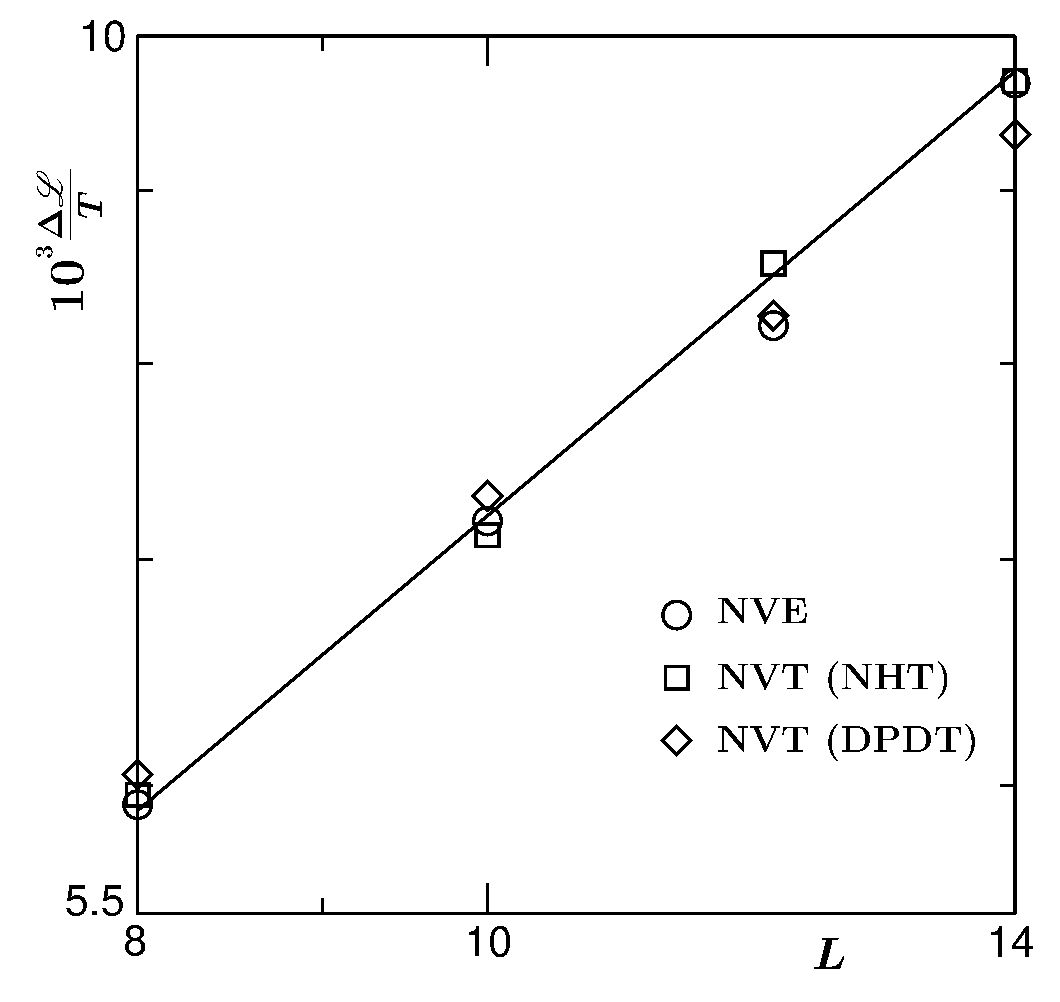}
\caption{\label{fig4}A finite-size scaling plot of Onsager coefficient $[{\mathscr L}={\mathscr L}
(\infty)]$, after subtracting the background contribution, using data at $T_c^L$s. Results 
from both NVE and NVT ensembles are shown. For NVT ensemble, we have included data from NHT and 
DPDT. The continuous line corresponds to the theoretical prediction for critical divergence.}
\end{figure}

\begin{figure}[htb]
\centering
\includegraphics*[width=0.4\textwidth]{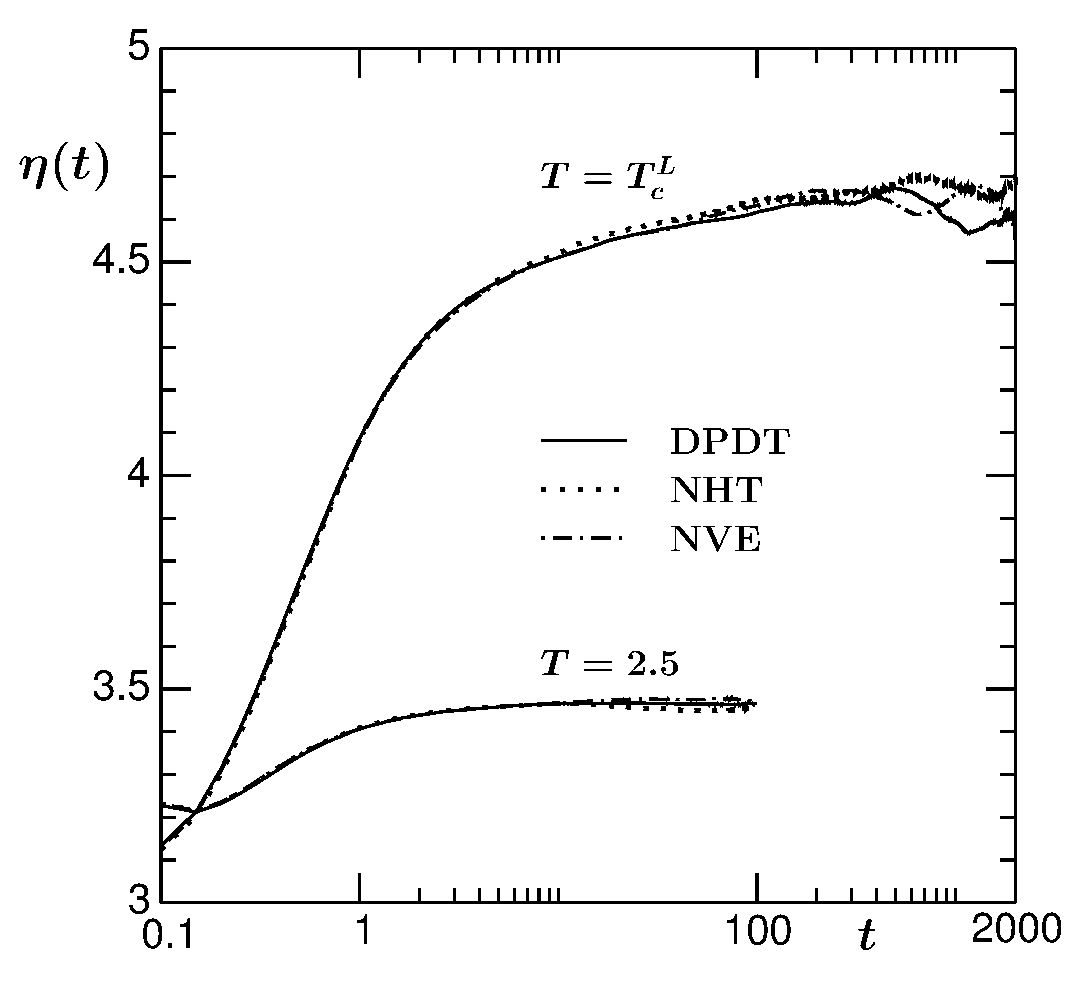}
\caption{\label{fig5}Plot of shear viscosity as a function of time for two different 
temperatures, viz., $T=2.5$ and $T=1.447$, the latter being the value of $T_c^L$ for 
$L=10$, the system size for which the results are presented. We have shown results from 
NVE, NHT and DPDT calculations.}
\end{figure}

\begin{figure}[htb]
\centering
\includegraphics*[width=0.4\textwidth]{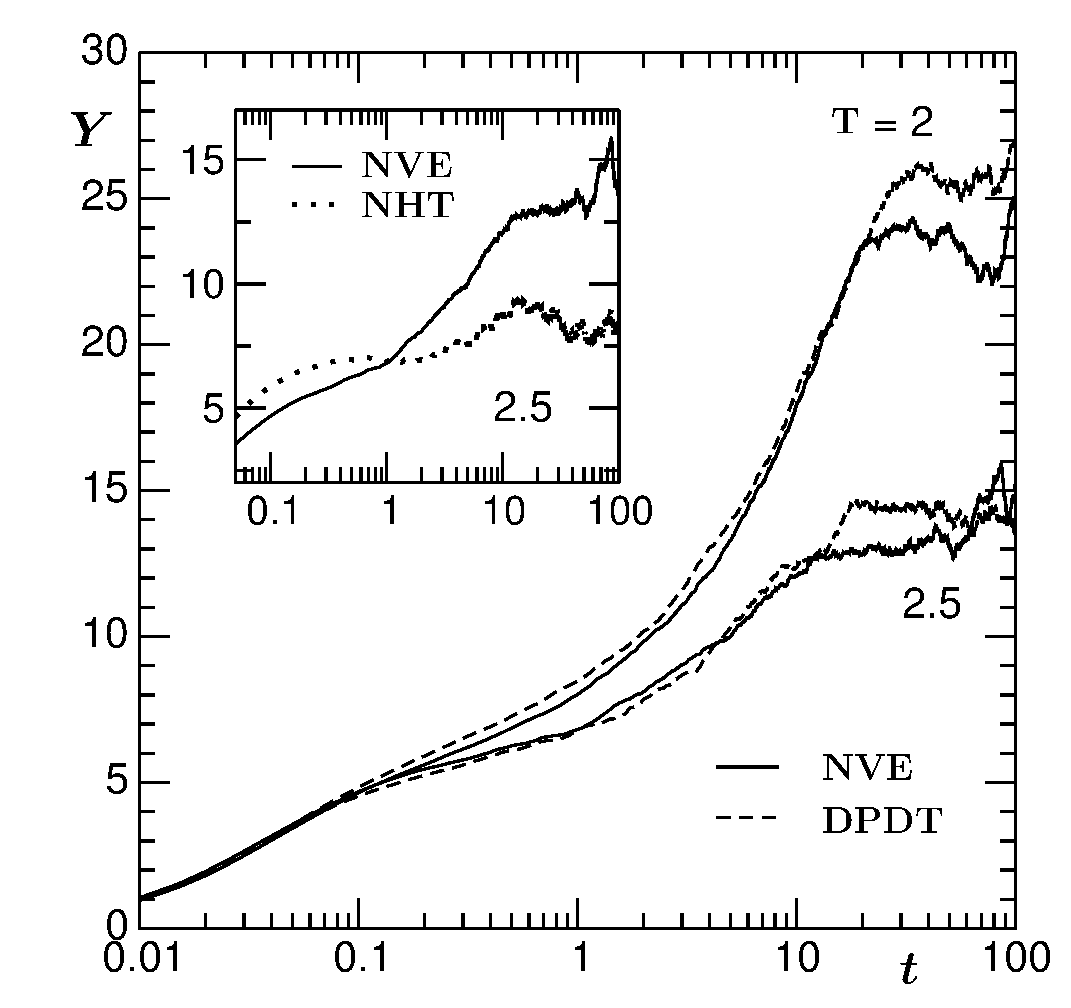}
\caption{\label{fig6}Comparison of time dependent bulk viscosity for calculations from 
NVE and DPDT, at two different temperatures. The inset shows a comparison between NVE and 
NHT for this quantity, only at $T=2.5$.}
\end{figure}

\par
\hspace{0.2cm}We start by showing a comparison of the time dependent self-diffusivity, 
calculated from the Einstein relation, in Fig. \ref{fig2}(a), obtained from NVE and NVT ensembles, 
at $T=2.5$. For NVT ensemble we have included results from AT, LT, NHT and DPDT as temperature 
controller. As expected, AT and LT do not provide results consistent with the NVE one. 
However, the results from NHT and DPDT are very much in agreement with the latter. The final 
values of the transport quantities, here and in other places, are obtained from the flat portions of these time-dependent 
plots. In Fig. \ref{fig2}(b) we show a comparison of $D$ calculated from NVE, NHT and DPDT, 
as a function of temperature, along the critical ($50:50$) composition line. All are in good 
agreement (the observed differences are not systematic). This is expected and demonstrated 
earlier \cite{frenkel}. However, the cases of collective properties (except for shear viscosity, via NHT, in a recent 
work \cite{roy2014}) are missing in the literature which we address below. 

\par
\hspace{0.2cm}In Fig. \ref{fig3} we show a comparison similar to Fig. \ref{fig2}(a) but for 
the time-dependent Onsager coefficient. For NHT, even though we have 
presented the result using only $Q=1$, we have performed the calculations with values of $Q$ 
up to $100$ and observed that the results are not very sensitive to the choice of this parameter. 
This fact will be demonstrated later, for all the transport quantities, by presenting representative 
results using the optimum value \cite{nose} of $Q$ ($\sim 6N k_B T/\omega_0^2$, $\omega_0$ being 
a characteristic vibrational frequency whose value is approximately $10$ for typical LJ fluid). 
Again, very good agreement is observed for results from NVE, NHT and DPDT. In the following we 
focus on the critical behavior of this quantity.

\par
\hspace{0.2cm}Note that $\mathscr L$ is expected to diverge at criticality 
with the correlation length $\xi$ as \cite{onuki1}
\begin{eqnarray}\label{cri4}
\frac{\mathscr L}{T}\sim \xi^{x_\lambda},
\end{eqnarray}
with $x_\lambda \simeq 0.9$. To verify the consistency of our simulation results with this number 
for the critical exponent, 
we take the route of finite-size scaling analysis \cite{fisher2}. Noting that at criticality 
$\xi$ scales with $L$, for results obtained at $T_c^L$, 
\begin{eqnarray}\label{cri5}
\frac{\mathscr L}{T} \sim L^{x_\lambda}.
\end{eqnarray}
It was observed in previous NVE MD simulations of this model \cite{das3,das4} that $\mathscr L$ 
has strong background contribution ${\mathscr L}_b$. The value of ${\mathscr L}_b$ was 
estimated to be $\simeq 0.0033$, a reasonably large number, given that for small system sizes 
this number can be comparable to the total value. 
We will thus deal with the critical 
part $\Delta {\mathscr L} (={\mathscr L}-{\mathscr L}_b)$ only. So, when calculated at $T_c^L$s, 
a plot of $\Delta {\mathscr L}/T$ vs $L$ will be consistent with a power-law with exponent $0.9$. 
This is demonstrated in Fig. \ref{fig4}. Note that we have shown results from NHT, DPDT, as well 
as from NVE ensemble. All of them are in good agreement. This essentially 
demonstrates that NHT and DPDT are good devices for the calculation of mutual diffusivity 
($D_{AB}$) even for quantitative understanding of critical dynamics. Here note that 
$D_{AB}={\mathscr L}/{\chi}$, where $\chi$ is the concentration susceptibility that can be 
conveniently calculated from concentration fluctuation in Monte Carlo simulations. Sightly poorer 
agreement of the DPDT data with the expected theoretical behavior, compared to NHT ones, is 
due to the temperature control problem that this method suffers from, in the long run.

\begin{figure}[htb]
\centering
\includegraphics*[width=0.4\textwidth]{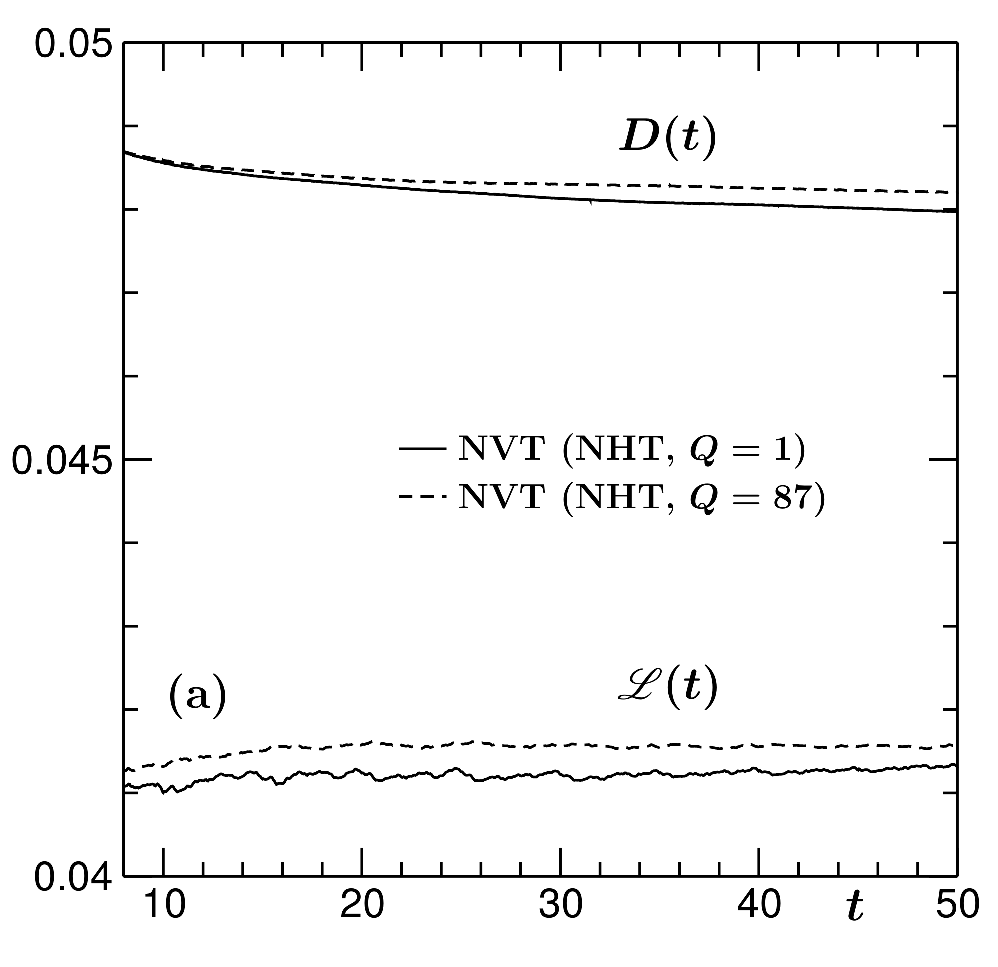}
\vskip 0.2cm
\includegraphics*[width=0.4\textwidth]{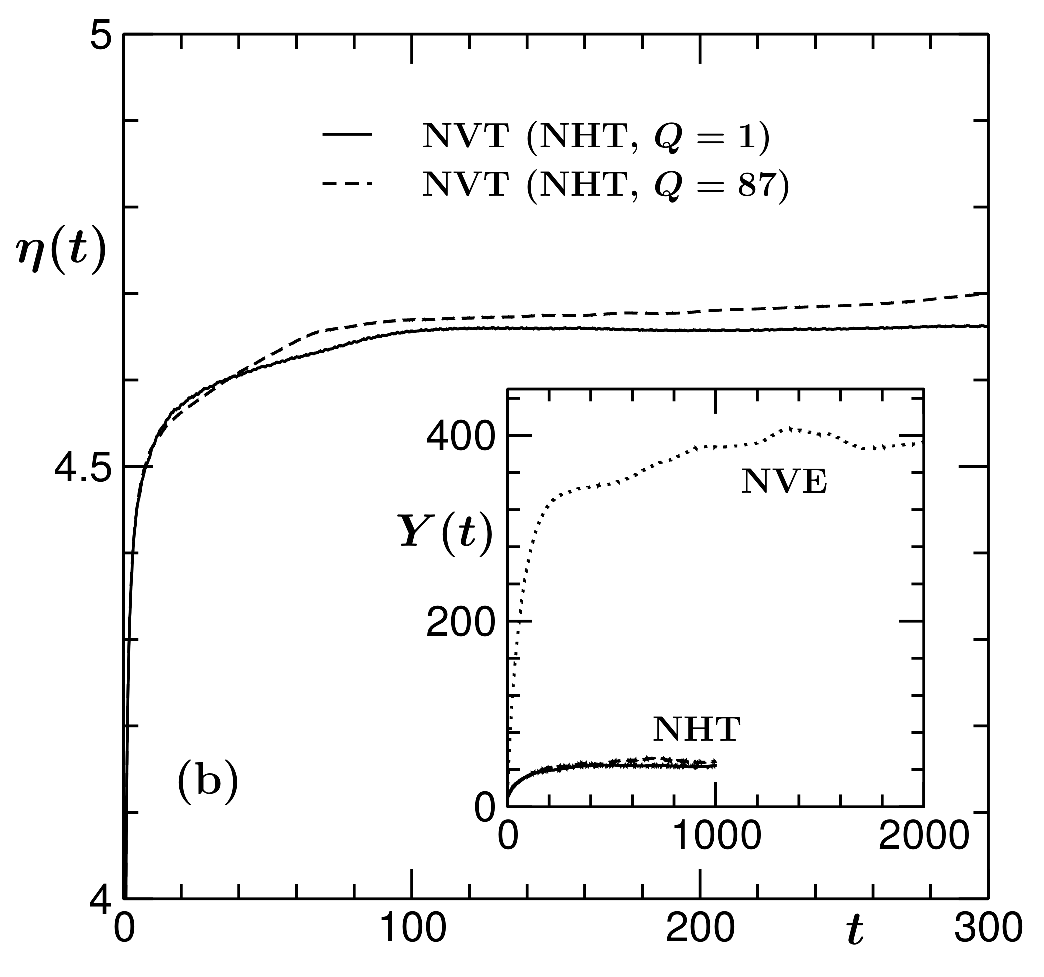}
\caption{\label{fig7}(a) A comparative plot of $D(t)$ and $\mathscr L(t)$, obtained using NHT, for two 
values of $Q$. Results correspond to $L=10$, $T=T_c^L$. Data for $\mathscr L$ has been multiplied by $3$. 
(b) Same as (a), but for $\eta(t)$ (main frame) and $Y(t)$ (inset).}
\end{figure}

\par
\hspace{0.2cm}Having demonstrated the usefulness of NHT and DPDT in the calculation of the diffusion constants, 
we turn our attention to viscosities. In Fig. \ref{fig5} we show the time dependent shear 
viscosity, using the Einstein relation, for NVE, NHT and DPDT. Two different temperatures are included, viz., 
$T=2.5$ and $T=1.447$. Here we do not show the results obtained 
using AT and LT which, we have already understood and as is known, are not appropriate for the study of 
transport properties in fluids. For both NHT and DPDT, satisfactory agreement is achieved with the NVE 
calculation. In our recent work \cite{roy2014}, agreement between the NHT and NVE was established. 
There our estimations of the corresponding critical exponent via these two methods agreed nicely with each other. 
However, in this work, DPDT 
was not applied. Having demonstrated the expected usefulness of DPDT, for this purpose, we move to the 
case of bulk viscosity. For bulk viscosity we avoid demonstrating the critical divergence, by 
keeping the difficulty in estimation of this quantity in mind. One of the primary difficulties lies in 
the estimation of $P$ that needs to be subtracted from the diagonal elements of the pressure 
tensor. Even a slight error in this quantity can lead to a misleading number in the final 
value. This, however, in our calculations was appropriately taken care of. Here note again that, for 
all the collective transport properties discussed in 
this work, the critical divergences were estimated from calculations via MD in NVE ensembles and the 
results are in good agreement with existing theoretical predictions. Due to the above mentioned 
difficulty and diverging relaxation time, it becomes inevitable to choose temperatures 
reasonably far away from the critical value, for the bulk viscosity.

\par
\hspace{0.2cm}Even though so far it appears that the NHT is a good tool to study dynamic critical 
phenomena, in fact better than the DPDT, from the temperature control point of view, 
we have encountered difficulty in the calculation of bulk viscosity, 
at least for this value of the coupling constant. In Fig. \ref{fig6} we show time-dependent
bulk viscosity. Good agreement (within $10\%$) is obtained between NVE and DPDT for two different 
temperatures. However, note that, possibly due to temperature fluctuation/drift, agreement between 
NVE and DPDT is not good if data from later parts of the runs are considered for the calculation. 
This is despite the fact that for this particular calculation we have used $\gamma=0.001$. 
Choice of a smaller value of $\gamma$ has connection with adopting smaller integration
time step. From a previous simulation \cite{nikunen}, it was reported that temperature destabilizes with
the increase of $\Delta t$. As
already mentioned, unlike other 
quantities, for $\zeta$, error in the calculation of $P$ brings additional 
problem, which enhances further if there is strong temperature fluctuation or drift. A 
further comparison of time dependent bulk viscosity is shown in the inset of Fig. {\ref{fig6}}, using data 
from NVE and NHT calculations. Clearly, NHT provides a misleading value. In fact there is disagreement 
between the two calculations starting from the very early time. 

\section{Conclusion}\label{conclusion}

\par
\hspace{0.2cm}In this paper we presented comparative results for transport properties in a binary fluid 
mixture obtained from molecular dynamics \cite{frenkel} calculations in microcanonical and 
canonical ensembles. The focus is on the collective properties. Even at criticality the Nos\'{e}-Hoover 
thermostat (NHT) and dissipative particle dynamics thermostat (DPDT)
provide results for diffusivities and shear viscosity 
that are in excellent agreement with the calculations in a microcanonical ensemble. However, while the  
DPDT appears to work well for bulk viscosity also, the NHT fails 
for this purpose. 

\par
\hspace{0.2cm}The importance of the paper lies in the following fact. Very close to the critical point, 
for big enough systems, one needs extended simulation runs for the calculation
of transport properties. In that case, for runs in microcanonical 
ensemble, it becomes difficult to avoid drift in temperature. Thus, the calculation of the 
transports in canonical 
ensemble may be of help. The NHT still being a very commonly used thermostat for the 
study of dynamics in the canonical ensemble, despite the criticisms about it, 
one needs to check its validity in situations as nontrivial as critical 
dynamics. It will be interesting to find out why the calculation for bulk viscosity via NHT is unreliable, 
despite the latter being a good one for other transport properties. 
One may argue, given that we have presented results only for $Q=1$,
if the value of $Q$ is appropriately chosen, the NHT results for bulk viscosity
may match the numbers obtained from microcanonical simulations.
In Fig. \ref{fig7}, we demonstrate, as stated earlier, that
improvements do not occur even when optimum value of $Q$ is chosen. In this figure,
we present results for all the transport properties, vs. time, calculated at $T_c^L$ for
$L=10$, using $Q=1$ and $87$, the latter number being approximately the optimum value for
this quantity. Within statistical fluctuations, the results from both the
values of $Q$ are in nice agreement with each other, for all the quantities. For bulk
viscosity, we have included a plot from calculations in NVE ensemble as well. This was done
due to the following fact. While for all the other quantities, either in this paper or elsewhere \cite{roy2014},
we have explored comparison between NHT and NVE results in the close vicinity of
the critical point, the same is missing for bulk viscosity. In Fig. \ref{fig6}, 
the temperatures were chosen to be
significantly higher than $T_c$, keeping the
inferior temperature controling ability of DPDT and other technical difficulties
in the calculation of bulk viscosity in mind.

\par
\hspace{0.2cm}A criticism about NHT is that \cite{stoyanov,schmid,yong}, if there is 
external force, there is problem with momentum conservation. Recently, such a problem is being taken care of 
\cite{stoyanov,schmid,yong} by introducing a further soft pair potential and relative velocities. Despite 
some deficiencies, even the basic NHT appears to provide a reasonable description of 
dynamics for a number of quantities, as seen here. Even for nonequilibrium dynamics we have observed 
\cite{roypre} recently that this thermostat produces expected results.

\par
\hspace{0.2cm}On the other hand, despite its better ability to preserve hydrodynamics,  
DPDT does not appear to be very suitable for studies of dynamic critical phenomena because of temperature 
control problem. In this context, a recent work
by Gross and Varnik \cite{gross} should be discussed. For studing dynamic critical phenomena, these authors
proposed a mesoscopic approach, based on the lattice Boltzmann method. In addition to accounting for
hydrodynamic transport, this approach keeps the temperature inherently constant.

\section*{Acknowledgement}\label{acknowledgement} 
SKD and SR acknowledge financial support from the Department of Science and Technology, India, via Grant 
No SR/S2/RJN-$13/2009$. SR is grateful to the Council of Scientific and Industrial Research, India, for 
their research fellowship. The authors acknowledge correspondence with C. Pastorino and T. Kreer, as well 
as useful comments from an anonymous referee. SKD 
acknowledges financial support from Marie-Curie Action Plan of European Commission (FP7-PEOPLE-2013-IRSES 
Grant No. $612707$, DIONICOS).

\vskip 0.5cm
\par
$*$~das@jncasr.ac.in

\end{document}